\documentclass[prd]{revtex4}

\usepackage{graphicx}
\usepackage{amssymb}
\usepackage{amsmath}

\newcommand{\be}{\begin{equation}}
\newcommand{\ee}{\end{equation}}
\newcommand{\ba}{\begin{eqnarray}}
\newcommand{\ea}{\end{eqnarray}}
\newcommand{\ban}{\begin{eqnarray*}}
\newcommand{\ean}{\end{eqnarray*}}
\newcommand \nn {\nonumber}

\begin{document}

\large

\setlength{\baselineskip}{17pt} 

\title{Heavy Quarks Embedded in Glasma}

\author{Margaret E. Carrington}
\affiliation{Department of Physics, Brandon University,
Brandon, Manitoba R7A 6A9, Canada\\
and Winnipeg Institute for Theoretical Physics, Winnipeg, Manitoba, Canada}

\author{Alina Czajka}
\affiliation{National Centre for Nuclear Research, ul. Pasteura 7,  PL-02-093  Warsaw, Poland}

\author{Stanis\l aw Mr\' owczy\' nski}
\affiliation{Institute of Physics, Jan Kochanowski University, ul. Uniwersytecka 7, PL-25-406 Kielce, Poland \\
and National Centre for Nuclear Research, ul. Pasteura 7,  PL-02-093 Warsaw, Poland}

\date{May 12, 2020}

\begin{abstract}

Heavy quarks, which are produced at the earliest stage of relativistic heavy-ion collisions, probe the entire history of the quark-gluon plasma that is created in the collision. Initially the plasma is populated with chromodynamic fields which can be treated as classical. We study the transport of heavy quarks across such a system, which is called {\it glasma}, using a Fokker-Planck equation where the quarks interact with long wavelength chromodynamic fields. We compute field correlators which are used to calculate the collision terms of the transport equation. Finally, the energy loss and momentum broadening of heavy quarks in the glasma are studied. Both of these quantities are sizable and strongly directionally dependent. 

\end{abstract}

\maketitle

Heavy quarks act as test probes of quark-gluon matter created in relativistic heavy-ion collisions, see e.g. the review \cite{Prino:2016cni}. Because of their large masses, heavy quarks are produced at the earliest stage of the collision through hard interactions of partons from incoming nuclei. They subsequently propagate through the surrounding medium and lose a significant fraction of their initial energy. Heavy quarks with sufficiently high transverse momenta test the entire history of the system. 

The medium produced in relativistic heavy-ion collisions evolves rapidly towards a locally equilibrated quark-gluon plasma which expands hydrodynamically and ultimately experiences a transition to a hadron gas. Final momentum spectra of heavy quarks are mostly shaped in the long-lasting equilibrium phase which is relatively well understood. Effects of the pre-equilibrium phase are usually entirely ignored, but  calculations recently performed in a framework of kinetic theory \cite{Das:2017dsh,Song:2019cqz} suggest that these effects are sizable. We are interested in the even earlier phase when the medium is not described in terms of quasi-particles, as in a kinetic theory, but rather as a system dominated by strong classical fields. It has been argued in a recent paper by one of us \cite{Mrowczynski:2017kso} that this transient phase significantly influences heavy-quark spectra. The plasma populated with chromodynamic fields appears to be opaque not only because of its high energy density. The energy loss and momentum broadening in such a medium are significantly bigger than in an equilibrium plasma of the same energy density \cite{Mrowczynski:2017kso}. 

Within the framework of the Color Glass Condensate (CGC) approach, see {\it e.g.} the review \cite{Gelis:2012ri}, color charges of partons confined in the colliding nuclei act as sources of long wavelength chromodynamic fields which can be treated classically because of their large occupation numbers. The non-equilibrium system from the early stage of the collision is called {\it glasma} \cite{Gelis:2012ri}. The transport properties of this system have been studied in a series of recent publications \cite{Ruggieri:2018rzi,Sun:2019fud,Liu:2019lac} where various configurations of glasma have been simulated numerically, and Wong equations of motions of heavy quarks interacting with chromodynamic fields have been solved numerically. Our objective is to develop an analytically tractable approach to the problem.

After light quarks and gluons have reached equilibrium, heavy quarks need extra time to adjust to the state of the plasma because of their large masses and correspondingly large relaxation times. Such a situation is naturally described in terms of a Fokker-Planck transport equation. This approach has been repeatedly applied to heavy quarks \cite{Moore:2004tg,Svetitsky:1987gq,vanHees:2004gq,Mustafa:2004dr}, and the Fokker-Planck equation of heavy quarks which interact with soft classical fields instead of plasma constituents was derived in Ref.~\cite{Mrowczynski:2017kso}.
Our aim is to obtain the collision terms of this Fokker-Planck equation within the CGC framework. These terms provide in turn the energy loss and momentum broadening of heavy quarks in a glasma.  We apply the method developed in \cite{Chen:2015wia} in which the fields  between the receding nuclei are expanded in powers of the proper time $\tau$. We take into account the first two terms of the expansion. Magnitudes of both the energy loss and momentum broadening are shown to be sizable, and strongly directionally dependent due to the glasma's anisotropy.   

The Fokker-Planck equation of heavy quarks embedded in a plasma system populated with strong chromodynamic fields is \cite{Mrowczynski:2017kso}
\ba
\label{F-K-eq}
\Big(D - \nabla_p^i  X^{ij}({\bf v}) \nabla_p^j - \nabla_p^i  Y^i({\bf v}) \Big) n(t, {\bf x},{\bf p}) = 0,
\ea
where
\ba
X^{ij}({\bf v}) &\equiv&
\frac{1}{2 N_c} \int_0^t dt'  \big\langle F_a^i(t, {\bf x}) F_a^j\big(t',{\bf x} - {\bf v}(t-t')\big)\big\rangle ,
\label{X-def}
\\[2mm]
\label{eq-X-vs-Y}
 Y^i({\bf v}) &=&
 X^{ij}({\bf v}) \frac{v^j}{T} .
\ea
$D \equiv\frac{\partial}{\partial t} + {\bf v} \cdot \nabla$ is the substantial derivative, the indices $i,\, j = 1, \, 2, \,3$ label the Cartesian coordinates $x,\, y, \, z$ and $\nabla_p$ is the momentum gradient. The color Lorentz force ${\bf F}_a(t,{\bf x})\equiv g \big({\bf E}_a(t,{\bf x}) + {\bf v} \times {\bf B}_a(t,{\bf x})\big)$ is expressed in terms of the chromoelectric ${\bf E}_a(t,{\bf x})$ and chromomagnetic ${\bf B}_a(t,{\bf x})$ fields, where $g$ is the QCD coupling constant. The quantities that carry color indices are written in the adjoint representation of the ${\rm SU}(N_c)$ gauge group with the indices  $a,\, b = 1, \, 2, \dots N_c^2 -1$. The notation $\langle \cdots \rangle$ denotes an ensemble average which assumes averaging over events in relativistic heavy-ion collisions. We use ${\bf v}$ for the heavy quark velocity, $E_{\bf p}$ is the energy of a heavy quark, and $n(t, {\bf x},{\bf p})$ is the distribution function of heavy quarks, which in equilibrium has the form $n^{\rm eq}({\bf p}) \sim \exp(- E_{\bf p}/T)$, where $T$ is the temperature of the equilibrated plasma of light quarks and gluons in which the heavy quarks are embedded. This equilibrium distribution should be a solution of the transport equation (\ref{F-K-eq}), which gives rise to the relation in Eq. (\ref{eq-X-vs-Y}). We use natural units with $c = \hbar = k_B =1$.

The collisional energy loss $dE/dx$ and transverse momentum broadening $\hat{q}$ of a heavy quark in the quark-gluon plasma can be obtained \cite{Mrowczynski:2017kso} from the relations
\be
\label{eloss-qhat-X}
\frac{dE}{dx}  =  - \frac{v}{T}  \frac{v^i v^j}{v^2} X^{ij}({\bf v}),
~~~~~~~
\hat{q}  =   \frac{2}{v} \Big(\delta^{ij} - \frac{v^i v^j}{v^2}\Big) X^{ij}({\bf v})  .
\ee

We next derive the correlators of the chromodynamic fields ${\bf E}$ and ${\bf B}$  that determine the tensor $X^{ij}({\bf v})$. 
We follow closely the presentation of \cite{Chen:2015wia}. The fields are generated by the color charges of the partons which are confined in the colliding nuclei during a relativistic heavy ion collision. The fields produced between the receding nuclei can be expanded in powers of the proper time $\tau$ as 
\be
\label{E-B-powers-tau}
\begin{array}{ll}
{\bf E} = {\bf E}_{(0)} + \tau \, {\bf E}_{(1)} + \tau^2 {\bf E}_{(2)} + \dots ,
\\[2mm]
{\bf B} = {\bf B}_{(0)} + \tau \, {\bf B}_{(1)} + \tau^2 {\bf E}_{(2)} + \dots . 
\end{array} 
\ee
The initial fields are parallel to the beam direction and are written ${\bf E}_{(0)} = (0,0,E^0)$ and  ${\bf B}_{(0)} = (0,0,B^0)$ with
\be
\label{E0-B0-adj}
\begin{array}{ll}
E^0_a({\bf x}_\perp) = -g f^{abc} A^i_{1b}({\bf x}_\perp) \, A^i_{2c}({\bf x}_\perp) ,
\\ [2mm]
B^0_a({\bf x}_\perp) = -g f^{abc}  \epsilon^{ij z} A^i_{1b}({\bf x}_\perp) \, A^j_{2c}({\bf x}_\perp) ,
\end{array} 
\ee
where $f^{abc}$ are the structure constants of the ${\rm SU}(N_c)$ group, $\epsilon^{ijk}$ is the totally asymmetric tensor, ${\bf x}_\perp =(x,y)$ is the transverse coordinate, and $A^i_{1a}({\bf x}_\perp)$ and $A^i_{2a}({\bf x}_\perp)$ are the pure gauge potentials generated by the incoming nuclei 1 and 2. The nucleus 1 moves with the speed of light in the positive direction of the collision axis $z$ and the nucleus 2 moves in the negative $z$-direction. The nuclei are homogeneous and infinitely extended in the transverse $x$-$y$ plane, and collide at $z=0$ and $t=0$. The potentials are purely transverse ${\bf A} = (A^x, A^y, 0)$ and they vanish beyond the forward light cone and therefore depend on the longitudinal coordinates $t$ and $z$ only through the step function $\Theta(t^2-z^2)$, which is not explicitly written. The first order fields are purely transverse and are
\be
\label{E1-B1-adj}
\begin{array}{ll}
\tau E_{(1)a}^i(t,{\bf x}_\perp,z) = - \frac{1}{2}\Big(
z \big(\partial^iE^0_a + g f^{abc} (A^i_{1b} +A^i_{2b})E^0_c \big)  
+ t \,\epsilon^{ijz} \big(\partial^jB^0_a + g f^{abc} (A^j_{1b} +A^j_{2b})B^0_c \big)  \Big) ,
\\[4mm]
\tau B_{(1)a}^i(t,{\bf x}_\perp,z) =  \frac{1}{2}\Big(
t \, \epsilon^{ijz} \big(\partial^jE^0_a + g f^{abc} (A^j_{1b} +A^j_{2b})E^0_c \big)  
- z \big(\partial^iB^0_a + g f^{abc} (A^i_{1b} +A^i_{2b})B^0_c \big)  \Big) .
\end{array} 
\ee

The correlators of the fields are determined by the correlators of the potentials generated by the same nucleus $\langle A^i_{1a} A^j_{1b} \rangle$ and $\langle A^i_{2a} A^j_{2b} \rangle$, since it is assumed that the potentials generated by the  different nuclei are uncorrelated: $\langle A^i_{1a} A^j_{2b} \rangle = 0$. The correlator of potentials generated by the same nucleus can be written
\be
\label{Ai0-Aj0-C1-C2}
\langle A^i_a({\bf x}_\perp) A^j_b({\bf x}'_\perp) \rangle 
= \delta^{ab}   \Big( \delta^{ij}_\perp  C_1 (r) - \hat{r}^i \hat{r}^j C_2 (r) \Big),
\ee
where $\delta^{ij}_\perp \equiv \delta^{ij} - \delta^{iz}\delta^{jz}$, ${\bf r} \equiv {\bf x}_\perp - {\bf x}'_\perp $, $r \equiv |{\bf r}|$, $\hat{r}^i \equiv r^i/r$.
We note that because the potentials are purely transverse and $r^z = 0$, the indices $i,j$ effectively run only through $x$ and $y$ in Eqs.~(\ref{E0-B0-adj}) - (\ref{Ai0-Aj0-C1-C2}). The functions $C_1(r),~C_2(r)$ are
\be
\label{C1-C2-def}
C_1(r) \equiv  \frac{m^2 K_0 (mr)}{g^2 N_c  f(r) }
\bigg[e^{\frac{g^4 N_c \mu f(r)  }{4\pi m^2 (N_c^2 -1)}} -1 \bigg] ,~~~~~~
C_2(r) \equiv  \frac{m^3 r \, K_1(mr) }{g^2 N_c  f(r) }
\bigg[ e^{\frac{g^4 N_c \mu f(r)  }{4\pi m^2 (N_c^2 -1)}} -1 \bigg] ,
\ee
where $f(r) \equiv \big( mr K_1(mr) - 1\big)$, $K_0(x)$ and $K_1(x)$ are the Macdonald functions and $m$ is an infrared regulator. Due to confinement color charges in a nucleus are neutralized at the length of a nucleon size which coincides with the inverse QCD scale parameter $\Lambda_{\rm QCD}$, and we therefore take $m \approx \Lambda_{\rm QCD} \approx  200$~MeV. The parameter $\mu$ is the charge density per unit transverse area of an incoming infinitely contracted nucleus and is expressed through the saturation momentum parameter $Q_s$ as \break $\mu = g^{-4}\,(N_c^2 -1) Q_s^2$ \cite{Chen:2015wia}. The function $C_1(r)$ logarithmically diverges as $r \to 0$. Since one expects that the correlation function (\ref{Ai0-Aj0-C1-C2}) is constant for $r \le Q_s^{-1}$ within the CGC approach, see e.g.  \cite{JalilianMarian:1996xn,Fujii:2008km}, the function $C_1(r)$ is assumed to be equal to $C_1(Q_s^{-1})$ for $r \le Q_s^{-1}$.

The zeroth order correlators are easily found to be
\be
\label{Ez-Bz-corr}
\begin{array}{ll}
\langle E^z_a({\bf x}_\perp) \, E^z_b({\bf x}'_\perp)\rangle 
=  g^2 N_c \delta^{ab} M_E(r) ,
\\[4mm]
\langle B^z_a({\bf x}_\perp) \, B^z_b({\bf x}'_\perp)\rangle 
=  g^2 N_c \delta^{ab} M_B(r), 
\\[4mm]
\langle E^z_a({\bf x}_\perp) \, B^z_b({\bf x}'_\perp)\rangle 
= 0 ,
\end{array} 
\ee
where
\be
\begin{array}{ll}
M_E(r) \equiv 2 C_1^2 (r) - 2 C_1 (r) \, C_2 (r) + C_2^2 (r) , 
\\[4mm]
M_B(r) \equiv 2 C_1^2 (r) - 2 C_1 (r) \, C_2 (r) .
\end{array} 
\ee

Using $\tau \equiv \sqrt{t^2 - z^2}$ and $\tau' \equiv \sqrt{t'^2 - z'^2}$ the first order correlators are computed as
\ba
\label{E0-E1-5} 
\tau' \langle E^0_a({\bf x}_\perp) \,  E_{(1)b}^i(t',{\bf x}'_\perp,z') \rangle 
&=& - \frac{g^2}{2} \, N_c \delta^{ab} \, \hat{r}^i 
z' \, M_E'(r) ,\nonumber
\\[2mm]
\label{E1-E0-5} 
\tau \langle E_{(1)a}^i(t,{\bf x}_\perp,z) \, E^0_b({\bf x}'_\perp) \rangle 
&=& \frac{g^2}{2} \, N_c \delta^{ab} \, \hat{r}^i 
z \, M'_E(r) ,\nonumber
\\[2mm]
\label{E0-B1-5}
\tau' \langle E^0_a({\bf x}_\perp) \,  B_{(1)b}^i(t',{\bf x}'_\perp,z') \rangle 
&=& \frac{g^2}{2} \, N_c \delta^{ab} \,  \epsilon^{ij} \hat{r}^j
t' \, M'_E(r),\nonumber
\\[2mm]
\label{B1-E0-5}
\tau \langle B_{(1)b}^i(t,{\bf x}_\perp,z) \, E^0_a({\bf x}'_\perp)  \rangle 
&=& - \frac{g^2}{2} \, N_c \delta^{ab} \,  \epsilon^{ij} \hat{r}^j
t\, M'_E(r),\nonumber
\\[2mm]
\label{B0-E1-5}
\tau' \langle B^0_a({\bf x}_\perp) \,  E_{(1)b}^i(t',{\bf x}'_\perp,z') \rangle 
&=& - \frac{g^2}{2} \, N_c \delta^{ab} \, \epsilon^{ij} \hat{r}^j \,
t'  \,M'_B(r) , \nonumber
\\[2mm]
\label{E1-B0-5}
\tau \langle E_{(1)b}^i(t,{\bf x}_\perp,z) \, B^0_a({\bf x}'_\perp) \rangle
&=& \frac{g^2}{2} \, N_c \delta^{ab} \, \epsilon^{ij} \hat{r}^j \,
t\, M'_B(r), \nonumber
\\[2mm]
\label{B0-B1-5}
\tau' \langle B^0_a({\bf x}_\perp) \,  B_{(1)b}^i(t',{\bf x}'_\perp,z') \rangle 
&=& - \frac{g^2}{2} \, N_c \delta^{ab} \, \hat{r}^i \,
z'  \, M'_B(r) , \nonumber 
\\ [2mm] \nn
\tau \langle B_{(1)b}^i(t,{\bf x}_\perp,z) \, B^0_a({\bf x}'_\perp) \rangle
&=& \frac{g^2}{2} \, N_c \delta^{ab} \, \hat{r}^i \,
z \, M'_B(r) . 
\ea

\begin{figure}[t]
\begin{minipage}{82mm}
\centering
\vspace{-2mm}
\includegraphics[scale=0.23]{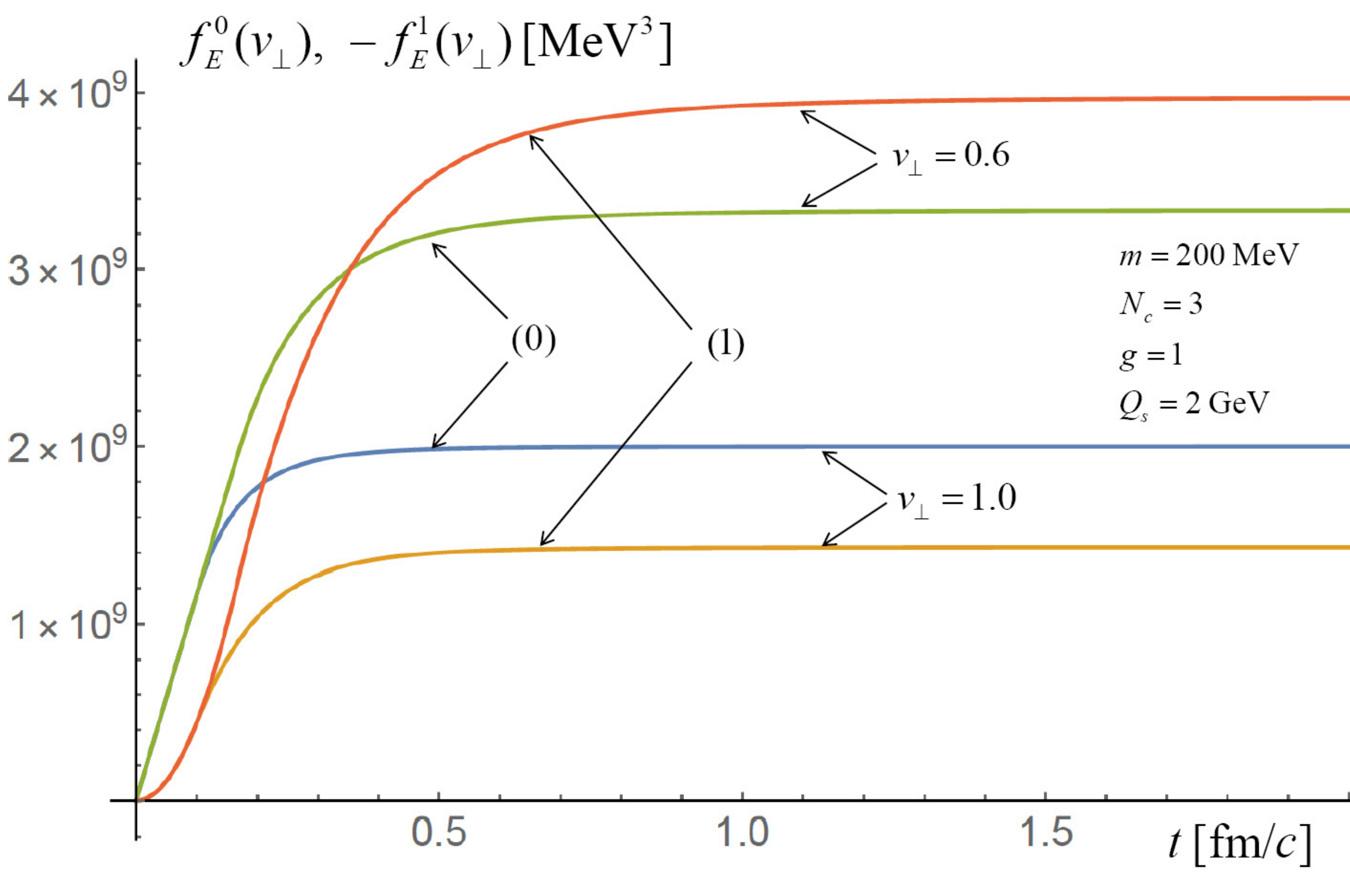}
\end{minipage}
\hspace{8mm}
\begin{minipage}{85mm}
\centering
\vspace{-2mm}
\includegraphics[scale=0.23]{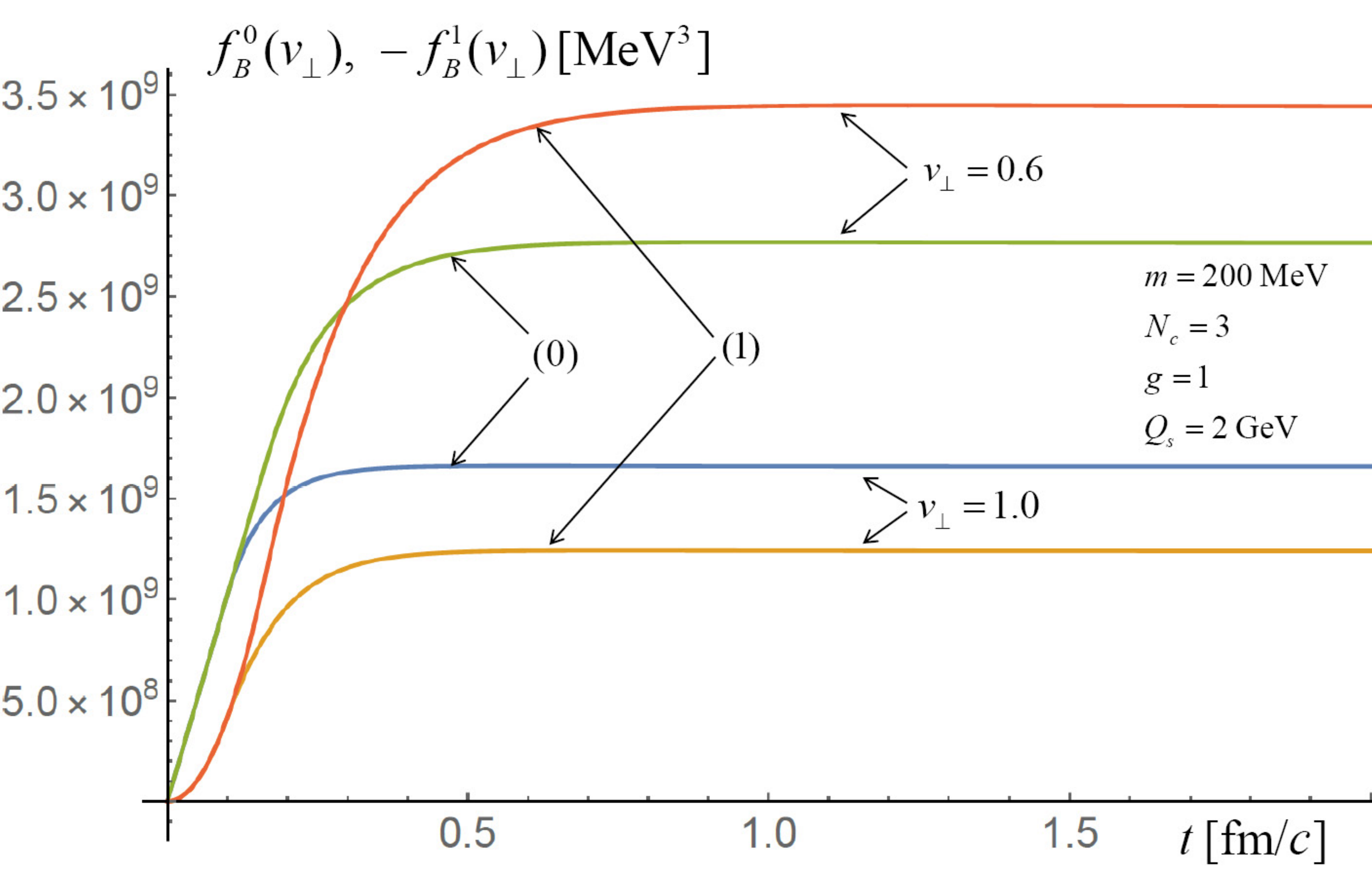}
\end{minipage}
\vspace{-2mm}
\caption{The pairs of functions ($f_E^0({\bf v}), -f_E^1({\bf v})$) and ($f_B^0({\bf v}),-f_B^1({\bf v})$) as functions of time $t$ for $v_\perp =0.6$ and $v_\perp =1$.}
\label{fig-fE-fB-01}
\end{figure}

Substituting the zeroth and first order correlators into Eq.~(\ref{X-def}) and using the notation where ${\bf x}' = {\bf x}-{\bf v} (t-t')$, $z' = z - v_\parallel(t-t')$, $r = v_\perp (t-t')$ and ${\bf n} = (0,0,1)$, we obtain
\ba
X^{ij}({\bf v}) &=&
\frac{g^2}{2 N_c} \int_0^t dt' \Big[\big\langle E_a^i(t, {\bf x}) E_a^j(t',{\bf x}')\big\rangle 
+ \epsilon^{jkl}v^k \big\langle E_a^i(t, {\bf x}) B_a^l(t',{\bf x}')\big\rangle \nn
\\[2mm]  
&&~~~~
+ \epsilon^{ikl}v^k \big\langle B_a^l(t, {\bf x}) E_a^i(t,{\bf x}')\big\rangle
+ \epsilon^{ikl} \epsilon^{jmn}v^k v^m \big\langle B_a^l(t, {\bf x}) B_a^n(t',{\bf x}')\big\rangle
\Big] 
\nn \\ [2mm]
&=&
\frac{g^4 (N_c^2-1)}{4} \int_0^t dt' \Big\{
2 n^i n^j M_E(r)
- \Big(n^i \hat{r}^j z'
- n^j \hat{r}^i z \Big) M'_E(r) \nn
\\[2mm]
&&~~~~ + \epsilon^{jkl}v^k \Big(
n^i n^n \epsilon^{lmn} \hat{r}^m t' + n^l n^n \epsilon^{imn} \hat{r}^m t \Big) M'_E(r)
\nn \\[2mm] 
&& 
~~~~ + \epsilon^{ikl}v^k \Big[
\epsilon^{jmn} v^m 
\Big( 2 n^l n^n M_B(r)
-  \big(n^l \hat{r}^n z'
-  n^n \hat{r}^l z \big) \, M'_B(r) \Big) \nn 
\\[2mm]
&& ~~~~ - \Big(n^l n^n \epsilon^{jmn} \hat{r}^m t' +  n^j n^n \epsilon^{lmn} \hat{r}^m t \Big) M'_B(r)
\Big] \Big\}. \label{X-corr-E-B}
\ea

The tensor (\ref{X-corr-E-B}) substituted into Eqs.~(\ref{eloss-qhat-X}) provides
\ba
\label{e-loss-f0-f1}
\frac{dE}{dx} &=& -\frac{v_\parallel^2}{vT} \Big[ f^0_E(v_\perp) + v_\perp f^1_E(v_\perp) \Big] ,
\\[4mm] 
\hat{q} &=& \frac{2 v_\perp^2}{v} \Big[ \frac{f^0_E(v_\perp)}{v^2} + f^0_B(v_\perp)
+ \frac{v_\perp}{v^2} f^1_E(v_\perp) 
+ \frac{(1 - v_\parallel^2)}{v_\perp} f^1_B(v_\perp)  \Big] ,
\label{qhat-f0-f1}
\ea
where
\ba
\label{fEB0-def} 
&& f^0_{E,B}(v_\perp) \equiv \frac{g^4 (N_c^2-1)}{2} \int_0^t dt' M_{E,B} (r) ,
\\[2mm]
&& \label{fEB1-def}
 f^1_{E,B}(v_\perp) \equiv \frac{g^4 (N_c^2-1)}{4} \int_0^t dt' (t-t') M'_{E,B}(r).~~~~~~
\ea

Our numerical results are obtained for $g =  1$, $N_c = 3$, $Q_s = 2$ GeV and $m=200$ MeV.  In Fig.~\ref{fig-fE-fB-01} we show how the pairs of functions $(f^0_E(v_\perp), -f^1_E(v_\perp))$ and $(f^0_B(v_\perp), -f^1_B(v_\perp))$ depend on time for $v_\perp = 0.6$ and $v_\perp = 1$. 
From the figure we see that for both velocities saturation is reached before $t=1.0$ fm/c. After saturation the corrections $-f^1_E(v_\perp)$ and $-f^1_B(v_\perp)$ are smaller than the zeroth order functions for $v_\perp =1.0$, but not for $v_\perp =0.6$. We have determined that the zeroth order contribution is larger for $v_\perp \gtrsim 0.73$. Thus we have that for large transverse velocities, saturation values are reached at times compatible with the small $\tau$ expansion introduced in equation (\ref{E-B-powers-tau}). 

\begin{figure}[t]
\begin{minipage}{86mm}
\centering
\vspace{-1mm}
\includegraphics[scale=0.275]{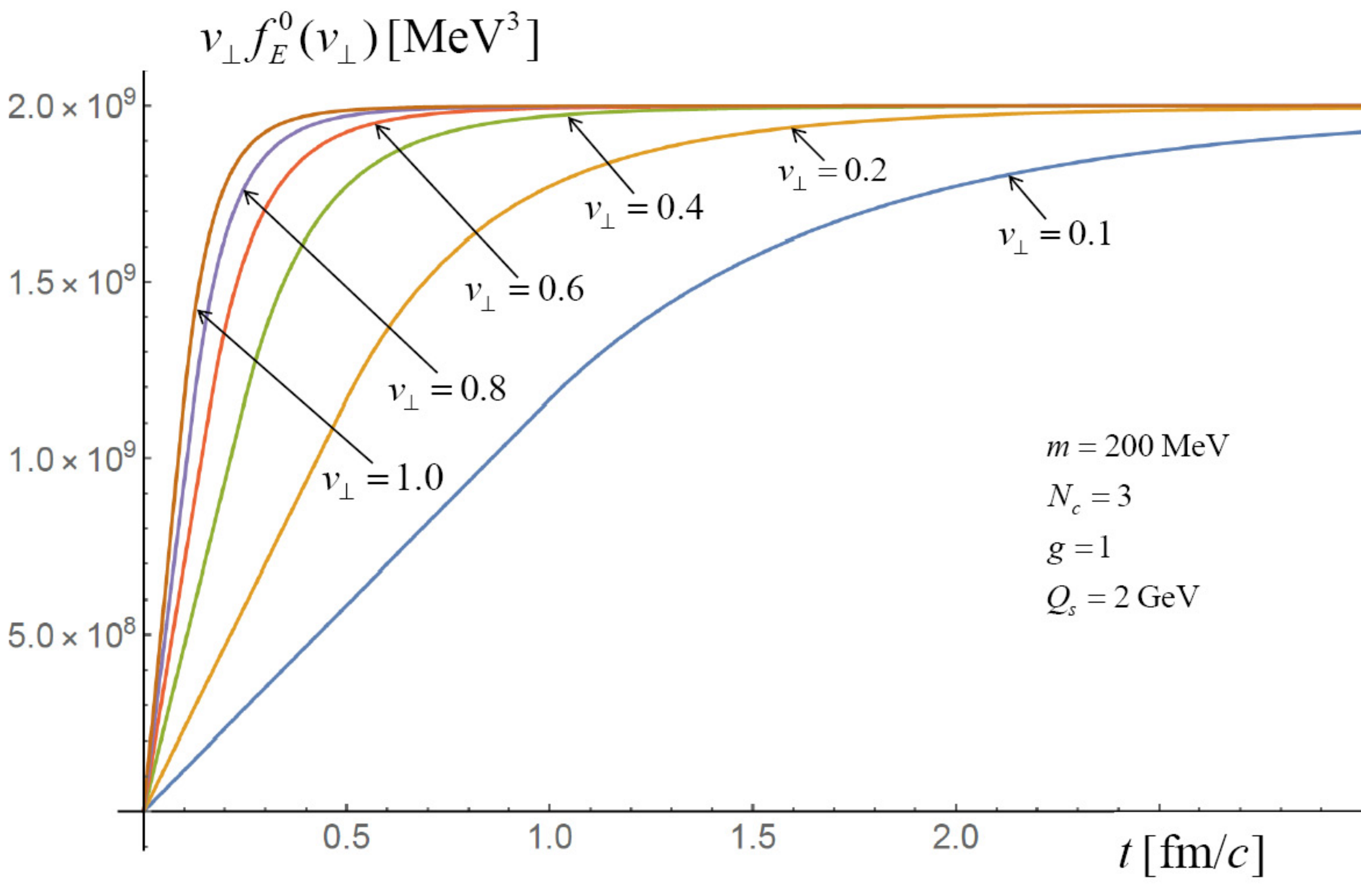}
\end{minipage}
\hspace{2mm}
\begin{minipage}{86mm}
\centering
\vspace{-5mm}
\includegraphics[scale=0.285]{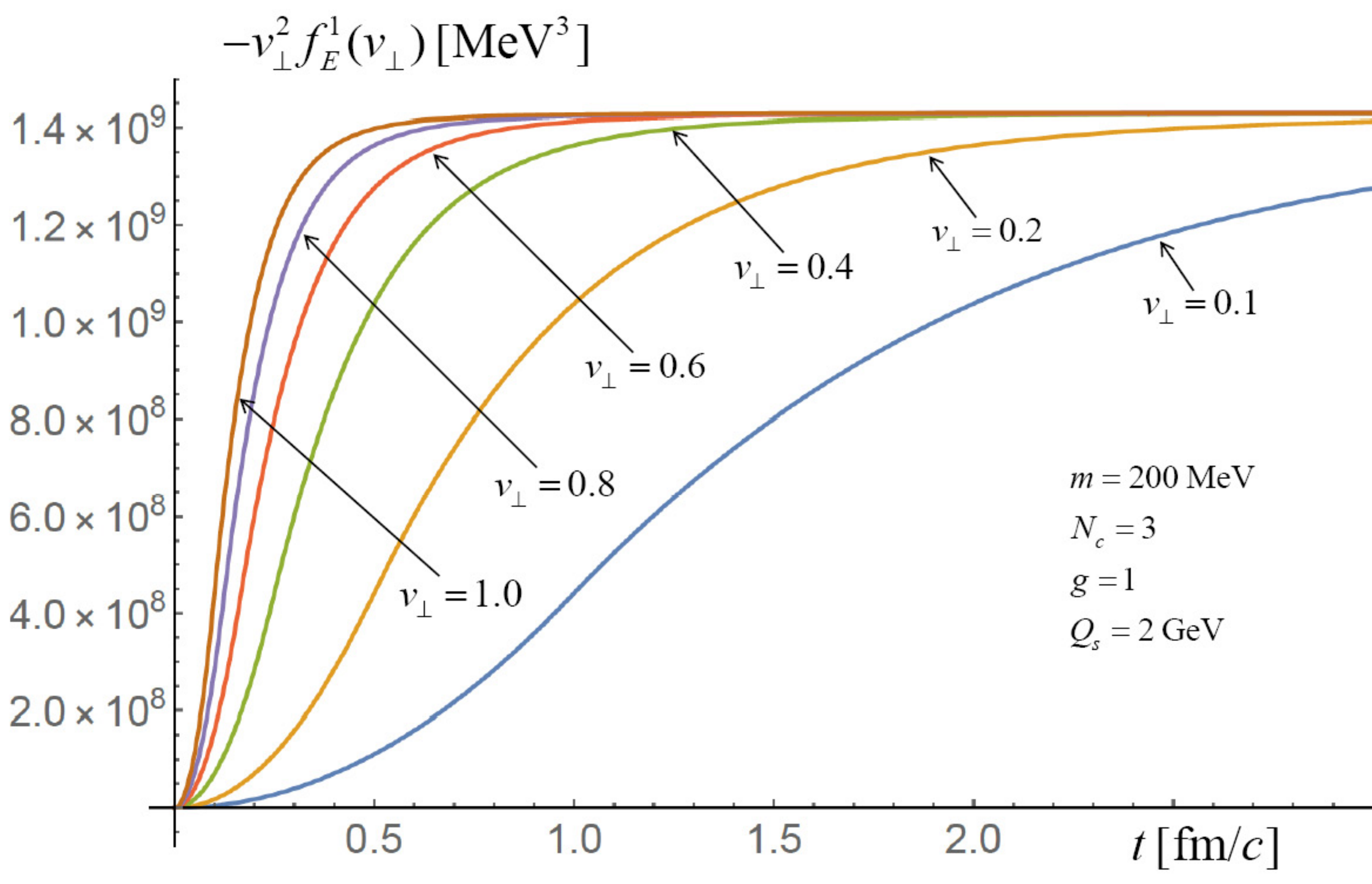}
\end{minipage}
\vspace{-1mm}
\caption{The quantities $v_\perp f_E^0(v_\perp)$ and $-v_\perp^2 f_E^1(v_\perp)$ as functions of time $t$ for six values of $v_\perp$.}
\label{fig-fE-fB-scaling}
\end{figure}

Since the integrands in the definitions (\ref{fEB0-def}) diverge at $r=0$, we have regularized them as
\be
\label{regularization}
M^{\rm reg}(r) \equiv \Theta(r_s-r) \, M (r_s) + \Theta(r-r_s) \, M (r) ,
\ee
with $r_s\equiv Q_s^{-1}$. Although the integrands in Eq.~(\ref{fEB1-def}) are regular at $r=0$, we have also regularized them according to the prescription (\ref{regularization}) because the dependence on $r$ for $r \le Q_s^{-1}$ is not physical in the CGC approach \cite{JalilianMarian:1996xn, Fujii:2008km}.
We have checked that our results are not strongly dependent on the regularization introduced in equation (\ref{regularization}).  
We observe that the values of all four functions saturate at sufficiently long times because the field correlators vanish for distances $r$ longer than a correlation length $r_c \sim m^{-1}$, which produces a saturation time $\sim r_c/v_\perp$. 

We also notice that the zeroth order functions $f_E^0(v_\perp)$ and $f_B^0(v_\perp)$ and the first order functions  $f_E^1(v_\perp)$ and $f_B^1(v_\perp)$ reveal interesting scalings. 
Changing the integration variable from $t'$ to $u=v_\perp(t- t')$ in the integrals (\ref{fEB0-def}) and (\ref{fEB1-def}), one shows that
if the integration extends over a big enough domain that the integrals saturate,
the  functions $f_E^0(v_\perp)$ and $f_B^0(v_\perp)$ depend on $v_\perp$ as $v_\perp^{-1}$ and  $f_E^1(v_\perp)$ and $f_B^1(v_\perp)$  as $v_\perp^{-2}$.  In Fig~\ref{fig-fE-fB-scaling} we show $v_\perp f_E^0(v_\perp)$ and $-v_\perp^2 f_E^1(v_\perp)$, which tend to universal values at large times. The behaviour of the functions $v_\perp f_B^0(v_\perp)$ and $-v_\perp^2 f_B^1(v_\perp)$ is similar.

\begin{figure}[t]
\begin{minipage}{88mm}
\centering
\includegraphics[scale=0.265]{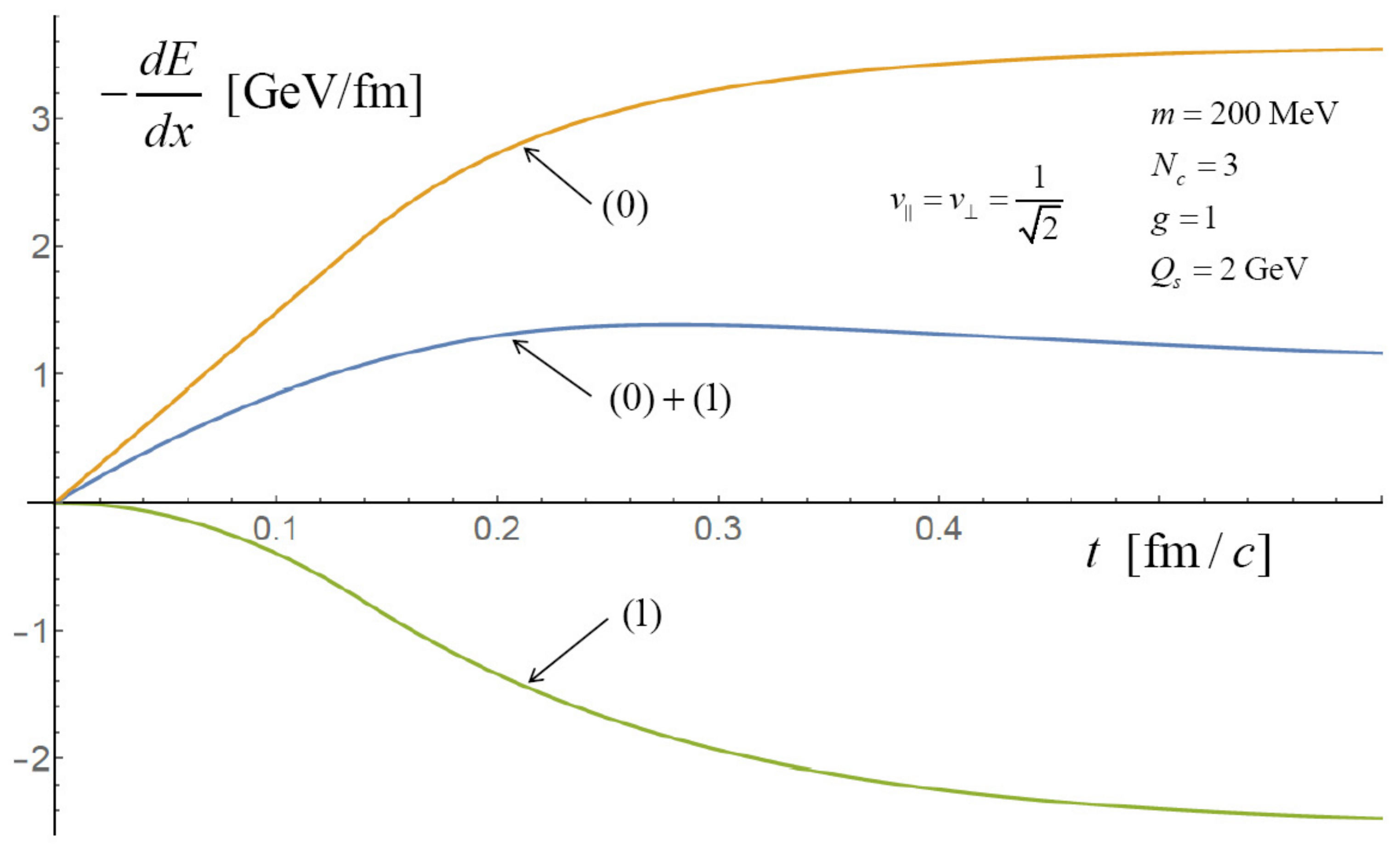}
\end{minipage}
\hspace{2mm}
\begin{minipage}{85mm}
\centering
\includegraphics[scale=0.265]{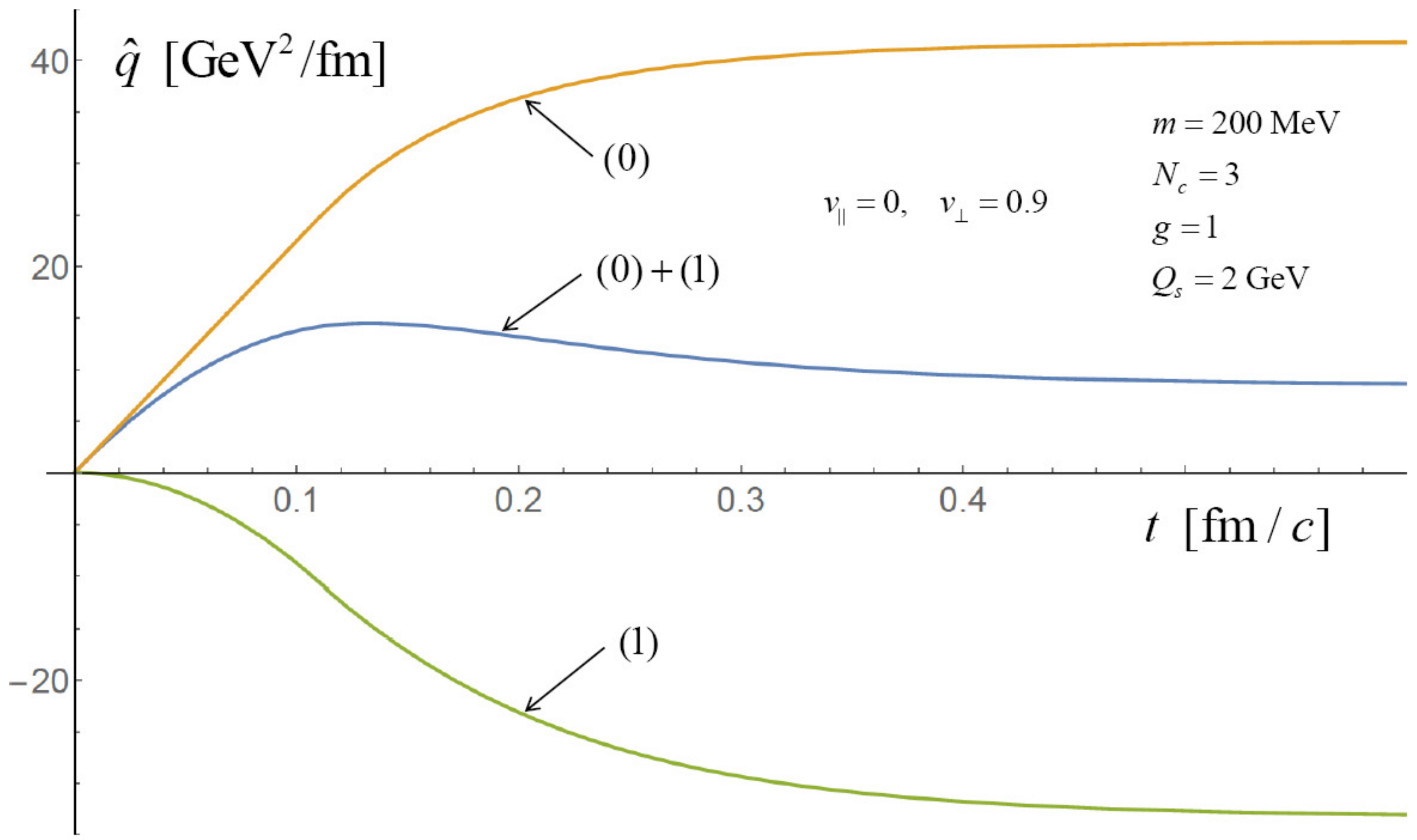}
\end{minipage}
\vspace{-1mm}
\caption{The energy loss $dE/dx$ for $v_\parallel = v_\perp = \sqrt{2}/2$ and the momentum broadening $\hat{q}$ for $v_\parallel = 0$ and $v_\perp = 0.9$ as functions of time $t$. The 0th and 1st order contributions and their sums are shown.}
\label{Fig-eloss-qhat-01}
\vspace{-4mm}
\end{figure}

In  Fig.~\ref{Fig-eloss-qhat-01} we present the energy loss for $v_\parallel = v_\perp = \sqrt{2}/2$ and the momentum broadening for $v_\parallel = 0$ and $v_\perp = 0.9$. Both quantities as functions of time first grow, reach a maximum, and then slowly decrease, which reflects the temporal evolution of the fields. 

\begin{figure}[t]
\begin{minipage}{82mm}
\centering
\vspace{-1mm}
\includegraphics[scale=0.23]{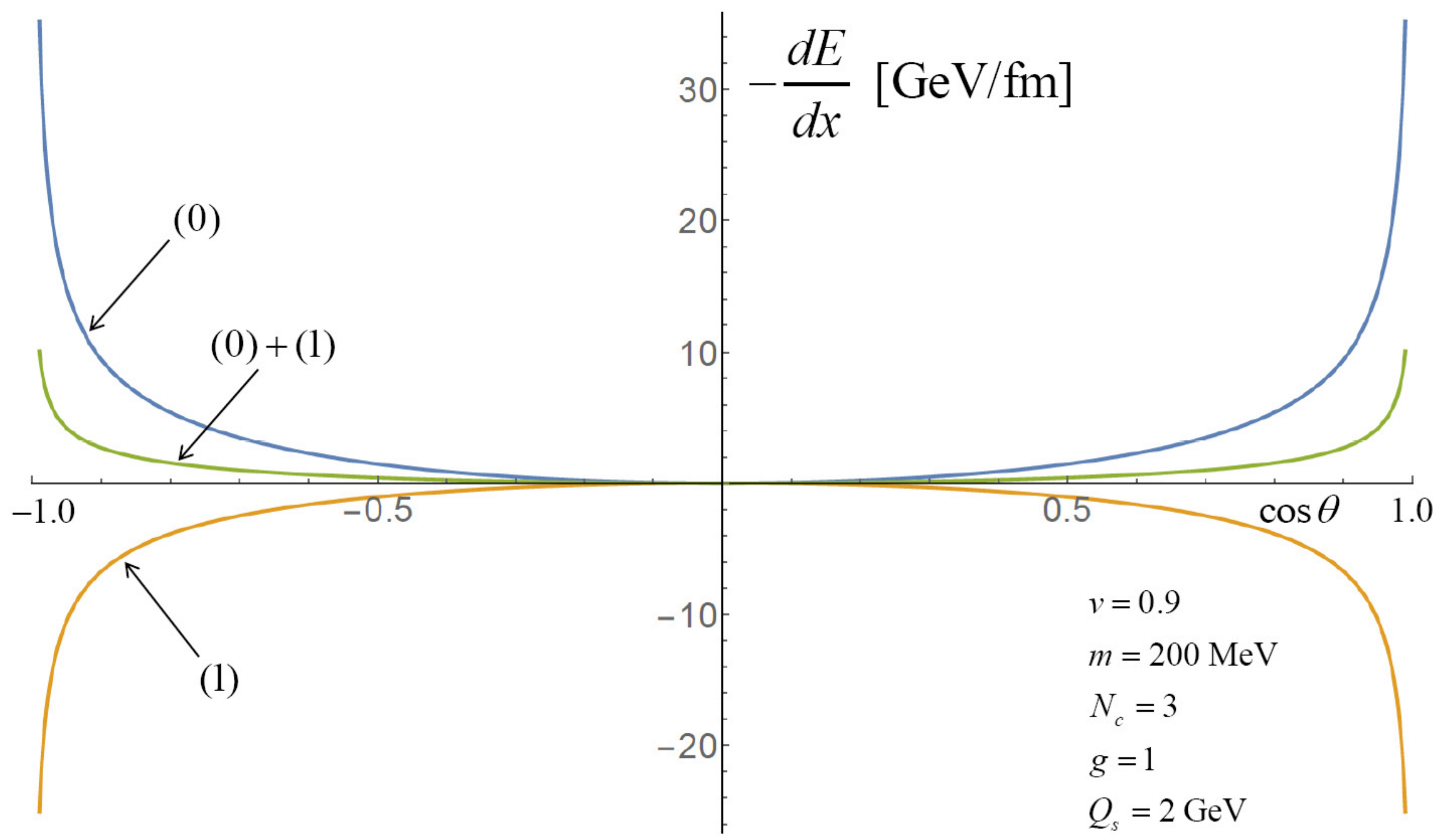}
\end{minipage}
\hspace{10mm}
\begin{minipage}{82mm}
\centering
\vspace{-1mm}
\includegraphics[scale=0.23]{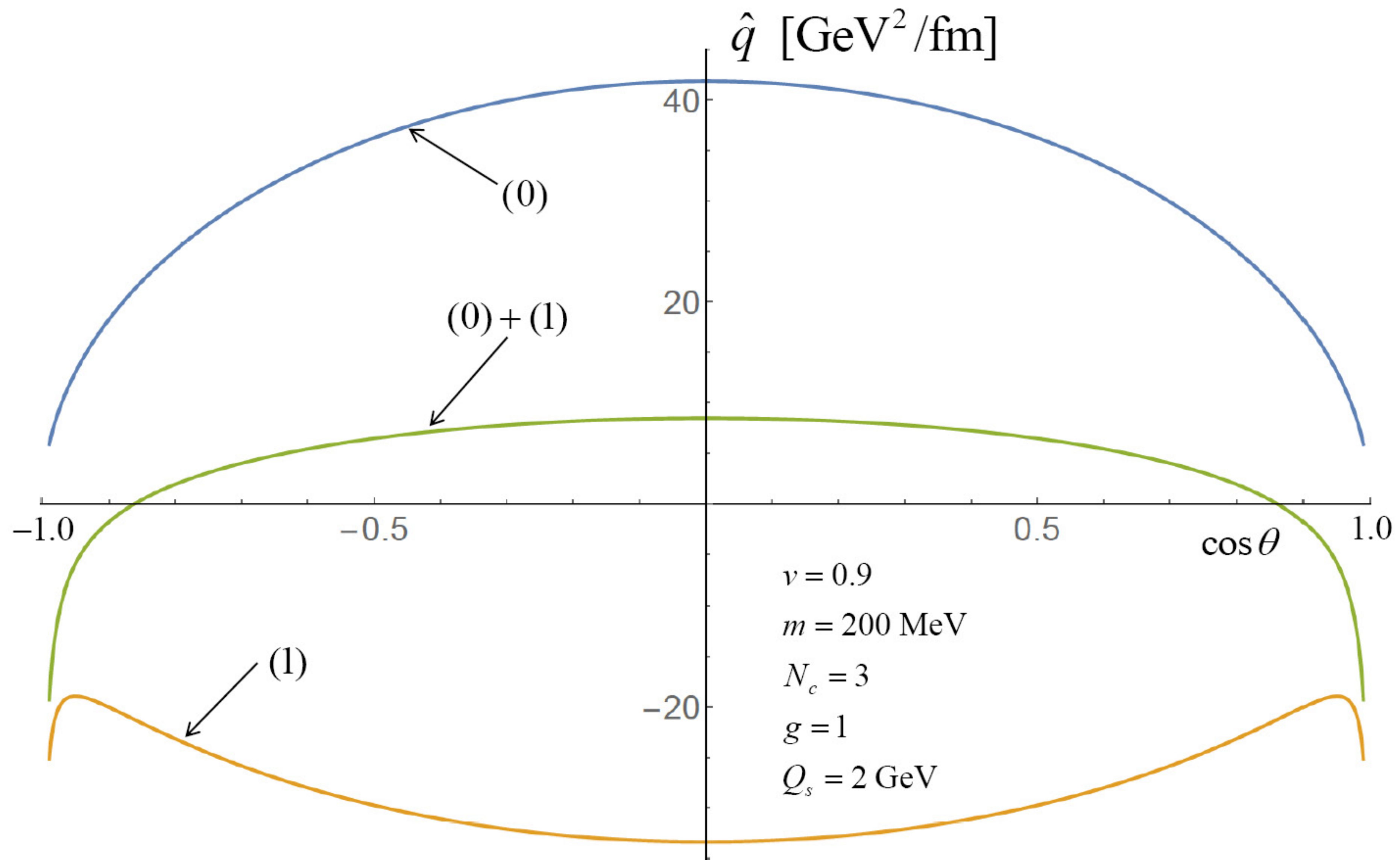}
\end{minipage}
\caption{The energy loss $dE/dx$  and the momentum broadening $\hat{q}$ as a function of $\cos\theta$ for $v=0.9$. We show the 0th and 1st order contributions to $dE/dx$ and $\hat{q}$ and their sums.}
\label{Fig-eloss-qhat-cos}
\end{figure}

In the left panel of Fig.~\ref{Fig-eloss-qhat-cos} we present the energy loss as a function of $\cos\theta$ where $\theta$ is the angle between the heavy-quark velocity and the collision axis $z$. The temperature $T$ is identified with the saturation scale $Q_s$ and the heavy-quark velocity is $v=0.9$. The time in the upper limit of the integrals in (\ref{fEB0-def}) and (\ref{fEB1-def}) is chosen large enough that $dE/dx$ reaches its saturation value for every $\cos\theta$. Because the chromoelectric field is mostly along the $z$-axis, the energy loss vanishes when a quark moves perpendicularly to this axis. The energy loss grows when the angle $\theta$ tends to 0 or $\pi$ and it becomes infinite for $\theta=0$ or $\theta=\pi$. The saturation time also becomes infinite when $v_\perp \to 0$ and consequently, as mentioned above, our results are not reliable in this limit. 

The momentum broadening (\ref{qhat-f0-f1}) is shown in the right panel of Fig.~\ref{Fig-eloss-qhat-cos}  as a function of $\cos\theta$. We see that $\hat q$ is maximal when a heavy quark moves perpendicularly to the collision axis and goes to zero when the angle $\theta$ tends  to 0 or $\pi$. When $\cos\theta$ approaches $\pm 1$ the magnitude of the first order contribution becomes bigger than the  zeroth order contribution. As explained previously, this signals that the small $\tau$ expansion (\ref{E-B-powers-tau}) is applicable only for the transverse velocities close to the speed of light.  

The numerical saturation value of $\hat{q}$ at $v=0.9$ and $\cos\theta=0$ equals $42~{\rm GeV^2/fm}$ when only 0th order contribution is taken into account. This value is similar to the result $70~{\rm GeV^2/fm}$ found in \cite{Mrowczynski:2017kso} using completely different reasoning, and is much bigger than the value of $\hat{q}$ inferred from a jet quenching in relativistic heavy-ion collisions which varies from $1.5$ to $7.0$  ${\rm GeV}^2/{\rm fm}$~\cite{Prino:2016cni}. When both the 0th and 1st order contributions to $\hat{q}$ are included, the value of $\hat q$ is reduced to $8.5~{\rm GeV^2/fm}$ which is much smaller than the zeroth order value, but still sizable. Our results suggest that in spite of its short lifetime the glasma can provide a significant contribution to  jet quenching. Higher order contributions need to be taken into account to draw a firm conclusion with regard to phenomenological consequences. 

In conclusion, we have derived the collision terms of the Fokker-Planck equation for heavy quarks embedded in a glasma, working up to first order in an expansion of the fields using the proper time $\tau$ as a small parameter. From the collision term, we have computed the energy loss and momentum broadening of heavy quarks in a glasma. The two quantities are strongly directionally dependent. The energy loss is maximal when a heavy quark moves along the collision axis and the momentum broadening has its maximum for a quark moving perpendicularly to the axis. The values of $dE/dx$ and $\hat{q}$ are sizable, suggesting that the glasma phase substantially contributes to the jet quenching observed in relativistic heavy-ion collisions. The zeroth and first order contributions to the collision term are of similar magnitude, which indicates that higher order terms could play an important role. The calculation of these higher order contributions, and a careful study of the effect of the regularization in Eq. (\ref{regularization}), is currently in progress. 

\vspace{3mm}

We are grateful to Rainer Fries for helpful correspondence. This work was partially supported by the National Science Centre, Poland under grant 2018/29/B/ST2/00646, and by the Natural Sciences and Engineering Research Council of Canada under grant  SAPIN-2017-00028.

\end{document}